\documentclass[aps,twocolumn,superscriptaddress,longbibliography]{revtex4-1}

\usepackage[colorlinks=true,citecolor=blue]{hyperref} 
\usepackage{graphicx}
\usepackage{bm}
\usepackage{amsmath}
\usepackage{amssymb}

\DeclareGraphicsExtensions{.png,.jpg,.eps}
\usepackage{xcolor}

\renewcommand{\H} {\mathcal{H}}
\newcommand{\rate} {\kappa/\kappa_{0}}

\begin{document}

\author{Wonhee Ko}
\affiliation{Center for Nanophase Materials Sciences, Oak Ridge National Laboratory, Oak Ridge, Tennessee 37831, USA}

\author{Jose L. Lado}
\affiliation{Department of Applied Physics, Aalto University, 02150 Espoo, Finland}

\author{Petro Maksymovych}
\email{maksymovychp@ornl.gov}
\affiliation{Center for Nanophase Materials Sciences, Oak Ridge National Laboratory, Oak Ridge, Tennessee 37831, USA}

\title{Non-contact Andreev reflection as a direct probe of superconductivity on the atomic scale}

\begin{abstract}
Direct detection of superconductivity has long been a key strength of point-contact Andreev reflection. However, its applicability to atomic-scale imaging is limited by the mechanical contact of the Andreev probe. To this end, we present a new method to probe Andreev reflection in a tunnel junction, leveraging tunneling spectroscopy and junction tunability to achieve quantitative detection of Andreev scattering. This method enables unambiguous assignment of superconducting origins of current-carrying excitations as well as detection of higher order Andreev processes in atomic-scale junctions. We furthermore revealed distinct sensitivity of Andreev reflection to natural defects, such as step edges, even in classical superconductors. The methodology opens a new path to nano- and atomic-scale imaging of superconducting properties, including disordered superconductors and proximity to phase transitions. 
\end{abstract}

\date{\today}

\onecolumngrid{ 
Notice: This manuscript has been authored by UT-Battelle, LLC, under Contract No. DE-AC0500OR22725 with the U.S. Department of Energy. The United States Government retains and the publisher, by accepting the article for publication, acknowledges that the United States Government retains a non-exclusive, paid-up, irrevocable, world-wide license to publish or reproduce the published form of this manuscript, or allow others to do so, for the United States Government purposes. The Department of Energy will provide public access to these results of federally sponsored research in accordance with the DOE Public Access Plan (http://energy.gov/downloads/doe-public-access-plan).
}

\newpage 
\clearpage 
\maketitle 

Andreev reflection (AR) occurs when electron traverses normal metal to superconductor interface, wherein a hole is retroreflected back into the metal and a Cooper pair is injected into the superconductor \cite{Andreev1964,Thikham2004}. By utilizing sharp metallic wire to make mesoscale contact with superconductors, point-contact AR (PCAR) has been developed as a very successful technique to probe the fundamental properties of superconductors \cite{Naidyuk2005,Daghero2013}. PCAR measurements are frequently interpreted with the Blonder-Tinkham-Klapwijk (BTK) formalism, which 
allows accounting both for BCS s-wave superconductivity\cite{Blonder1982}, unconventional and multiband superconductivity\cite{Kashiwaya1996,Brinkman2002}. 
While conventional Andreev reflection is currently well understood, atomic-scale tunneling techniques exploiting this phenomena for sensing the nature of substrate carriers have remained relatively unexplored. 

Scaling down Andreev measurements to nano- and even atomic-scales would provide a definitive methodology to directly probe microscopic properties of superconductivity, such as effects of disorder \cite{Dubouchet2018}, competing order parameters \cite{Park2008}, interfaces \cite{Oh2021} and topological defects \cite{Tanaka2009,Zhu2020}, extending
and complementing nanoscale measurements of quasiparticle density of states in scanning tunneling microscopy (STM) \cite{Pan1998,Hoffman2002} and break junctions \cite{vanderPost1994,Scheer1998}. However, bringing point contact measurements such as PCAR to atomic-scale measurements faces distinct challenges. On the one hand, atomic contacts are mostly random, invasive (causing local surface damage) and poorly controlled \cite{Tartaglini2013}, likely even more so for complex surfaces of unconventional superconductors such as cuprates \cite{Wei1998} and iron-based superconductors \cite{Daghero2011}. Meanwhile, the theoretical models would have to be extended to account for the atomic-scale contact geometry, unconventional order parameters and multiband gap structures\cite{Daghero2013}.

\begin{figure}[t!]
\center
\includegraphics[width=\linewidth]{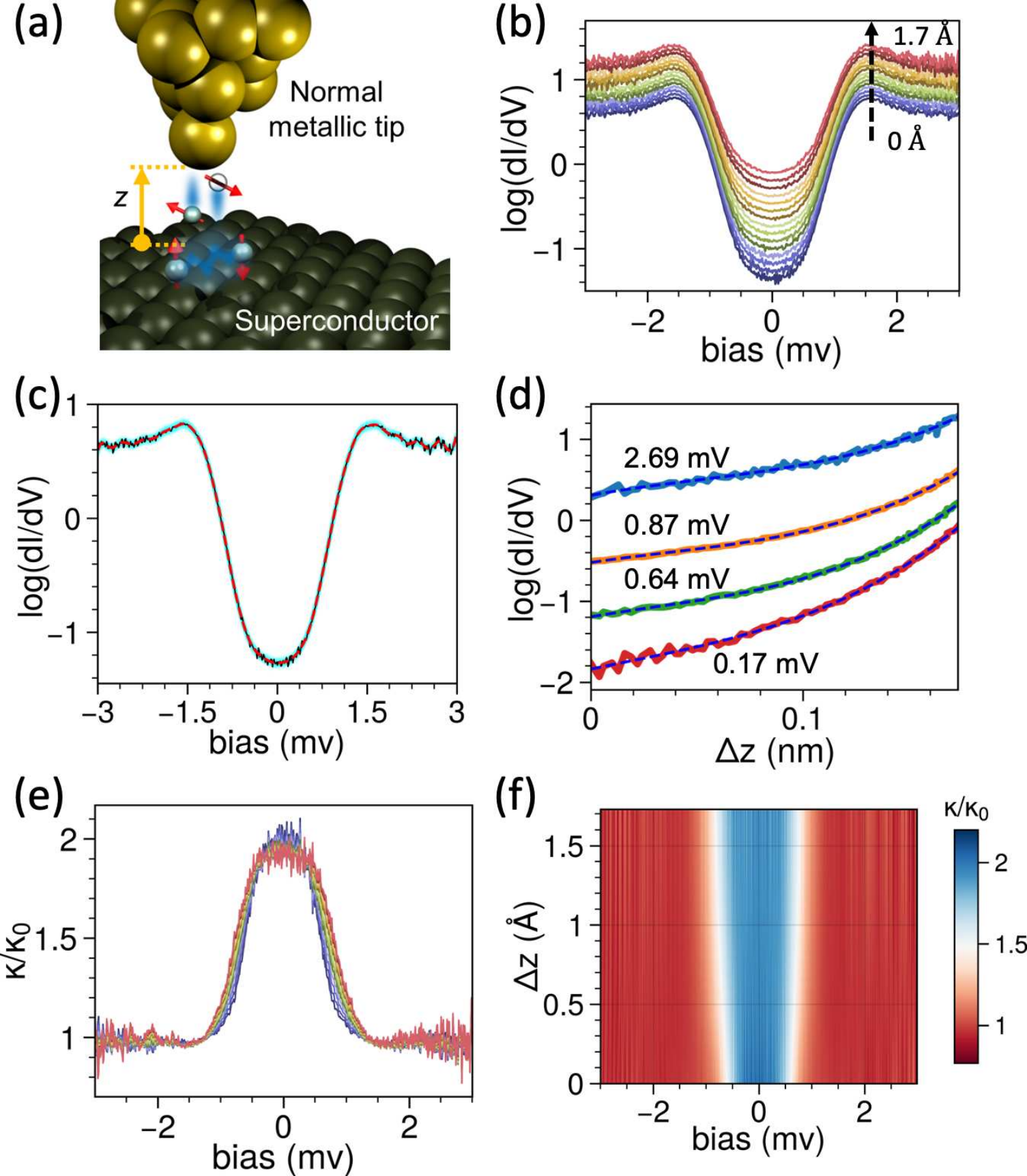}
\caption{(a) Schematic of the non-contact Andreev reflection measurement. (b) Differential conductance $dI/dV$ in log scale at different distances. (c) Raw $\log(dI/dV)$ spectra from numerical differentiation (cyan) and the gaussian process regression of the spectra for noise reduction (black). (d) Tip-surface ($\Delta z$) dependence of denoised $\log(dI/dV)$ for specific bias (as shown), with 2nd degree polynomial fit to each curve (dotted line). (e) Normalized decay rate $\rate$ at each $z$ extracted from the differentiation of $\log(dI/dV)$-$z$ curves. (f) Colormap plot of $\rate$ as a function of bias and $\Delta z$. z = 0 is a reference point corresponds to smallest tunneling conductance. }
\label{fig:fig1}
\end{figure}

In this letter, we develop a non-contact Andreev reflection (NCAR) technique based on scanning tunneling microscopy/spectroscopy (STM/S). The technique contributes tunability of junction geometry and tunneling spectroscopy to the measurement of Andreev reflection. However, it does so at the expense of direct measurement of contact conductance, which is the fundamental observable in PCAR \cite{Naidyuk2005}. To address this apparent challenge we exploit the relative probability of tunneling between normal and Andreev process as a robust observable sensitive to local Andreev reflection \cite{PhysRevResearch.3.033248}. The relative probability is quantified by measuring normalized decay rate of tunneling current with low-temperature STM, combined with detailed tight-binding modeling. The method emerges as an effective and direct probe of superconductivity up to the superconducting transition temperature ($T_c$). Moreover, our methodology directly reveals the order of the tunneling process responsible of the current, and the inhomogeneity of Andreev reflection due to single atomic steps even in classical superconductors. This combination of non-contact and near-contact Andreev reflection (NeCAR) enriches the spectrum of properties accessible with nano and atomic-scale resolution, enabling new insight into local properties of both classical and exotic superconductors.

\begin{figure}[t!]
\center
\includegraphics[width=\linewidth]{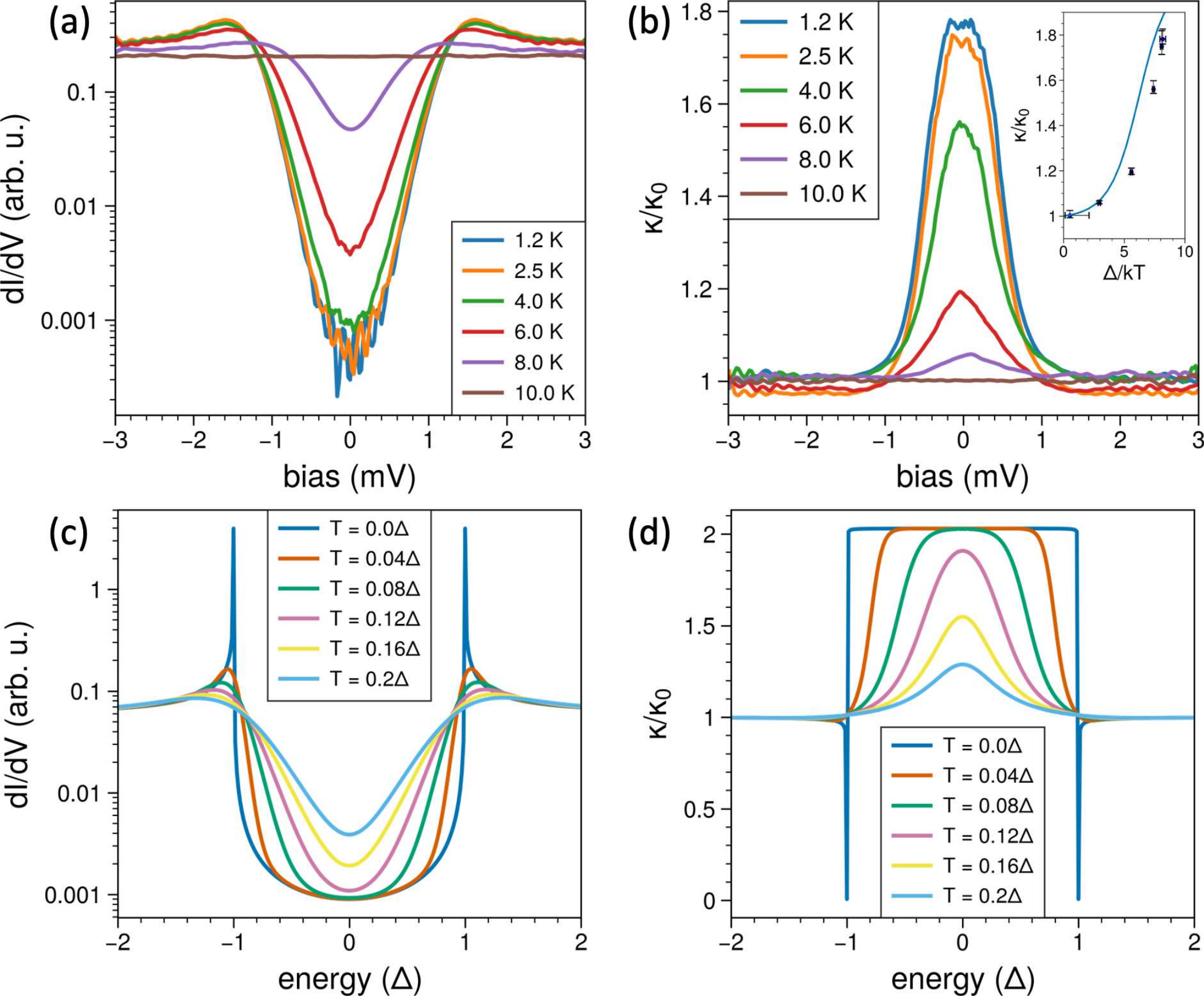}
\caption{Temperature dependence of non-contact Andreev reflection from experimental measurements (a,b) and theoretical (c,d) calculations. Thermal broadening causes progressive closing of the superconducting gap as seen in tunneling conductance $dI/dV$ (a,c), and a closely correlated reduction of the fraction of Andreev reflection as seen by reduction of $\rate$ (b,d). Inset of (b) shows the theoretical (blue line) and experimental (black dots) evolution of $\rate$ with respect to the ratio of superconducting gap and temperature.}
\label{fig:fig2}
\label{fig2}
\end{figure}

To reveal the properties of NCAR, we carried out detailed spectroscopy of  normal metal - superconductor tunnel junction in well-understood Pb(110) as a model system [Fig. \ref{fig:fig1}(a)]. The atomically clean surface of Pb(110) was prepared by repeating sputtering-annealing cycles \cite{Ruby2015}.
The experiments were performed in SPECS JT-STM operated at ultrahigh vacuum condition ($< 10^{–10}$ mbar) and base operating temperature of 1.2 K. The quality of STM tip was checked on either Au(111) or Cu(111) single crystals, and further conditioned by pulsing and soft crashes until the tip displayed proper metallic behavior. The $I$-$V$ curves were acquired by sweeping direct current bias without any additional electrical signals. The differential conductance $dI/dV$ were obtained by numerical differentiation of the $I$-$V$ curves. To minimize the noise in numerical differentiation, the analysis of the decay rate was carried out by fitting numerically obtained $\log(dI/dV)$ with Gaussian process regression [Fig. \ref{fig:fig1}(b)]\cite{scikit-learn}, which provides effective interpolation as well as a robust noise-level estimation. We utilized a simple kernel that is a sum of squared-exponential and white noise. We found the performance of this algorithm to be robust toward the choice of initial values and kernel parameters.

\begin{figure*}[t!]
\center
\includegraphics[width=0.8\linewidth]{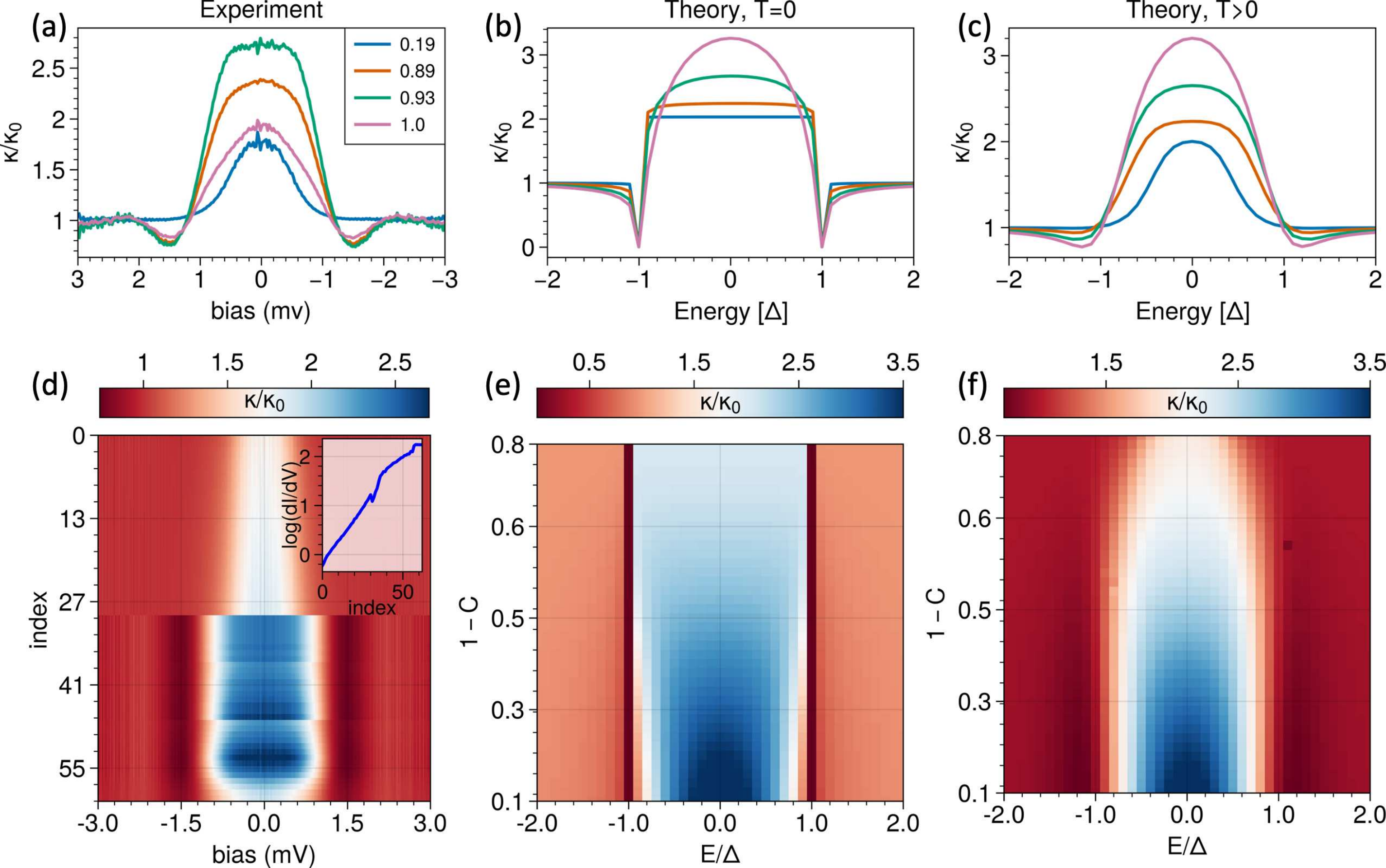}
\caption{Progression of Andreev reflection from non-contact to contact, revealing a unique near-contact regime. (a,d) The evolution of $\rate$ as a function of relative conductance from experiment (a) and theoretical calculations at T=0 (b) and T > 0 (c). (d-f) show more detailed dependence on conductance as a colormap of $\rate$. The discontinuities of $\rate$ stem from a mechanical deformation of the tip (Inset of (d)). In panels (e,f), y axis is a function of $C = v_0(t_S t_T)^{-1/2} e^{-\lambda z}$ that indicates the transparency of the junction, where $C=1$ the transparent contact regime.}
\label{fig:fig3}
\label{fig3}
\end{figure*}

To detect Andreev reflection, $dI/dV$ spectra were measured as a function of the distance between the tip and the sample [Fig. \ref{fig:fig1}(c)]. $dI/dV$ increases nearly exponentially both inside and outside of the superconducting gap. Exact height dependence was extracted from $dI/dV$-$z$ curves at each bias [Fig. \ref{fig:fig1}(d)], which can be well fit with an exponential function $(dI/dV)_{0}\cdot e^{-\kappa z}$ in tunneling regime (i.e. at tunneling conductance $G\ll G_\text{contact}\sim 2e^2/h$). One prominent feature is that although the amplitude $(dI/dV)_{0}$ changes, its slope in $\log(dI/dV)$-$z$ curve, i.e. decay rate $\kappa$, is almost constant outside of the superconducting gap. However, for the bias inside the gap, the slope significantly increases, which reflects a change of the dominant tunneling mechanism (conceptually similar to our previous approach to differentiate between Andreev and Josephson currents \cite{PhysRevResearch.3.033248}). Because of the normal-superconducting junction in the present case, the lowest order contribution to change in the tunneling decay rate is the onset of Andreev reflection.

The change in $\kappa$ was further quantified by calculating its dependence on bias from the $\log(dI/dV)$ fitted by Gaussian process regression, and subsequently normalizing by the decay rate of normal tunneling $\kappa_{0}$ measured well outside the superconducting gap. This procedure should account for the variation of the decay rate in the normal tunneling regime, such as geometrical and chemical structure of the tunnel junction. The resulting normalized decay rate $\kappa/\kappa_{0}$ should now be primarily sensitive to the superconducting state. As seen in Fig. \ref{fig:fig1}(f), $\kappa/\kappa_{0}$ maintains approximately constant value of 1 outside the gap, and it increases to a value of approximately 2 inside the gap. These are exactly the values expected for transition from single charge tunneling to Andreev reflection \cite{Cuevas1996,Cuevas2003,Johansson2003}. Moreover, $\kappa/\kappa_{0}$ is only weakly dependent on the specific value of tunneling conductance (set by tip-surface height) in the tunneling regime. As the tip brought closer to the surface, only the bias range where $\kappa/\kappa_{0}> 1$ becomes wider, which is displayed more clearly in the colormap plot of $\kappa/\kappa_{0}$ as a function of bias and $z$ [Fig. \ref{fig:fig1}(e)]. The observation indicates that the Andreev reflection becomes more prominent at the gap edge as the tunneling barrier gets narrower.

Several factors can induce the deviation from ideal value of $\kappa/\kappa_{0}=2$ of Andreev reflection, most prominently instrumental and thermal broadening. Figure \ref{fig2}(a) and \ref{fig2}(b) shows the $dI/dV$ spectra and normalized decay rate, respectively, as temperature increased from 1.2 K to 10 K. The superconducting gap in $dI/dV$ displays both thermal broadening and gap closing as temperature is increased, until there is no gap as $T > T_c$ at 10 K. Meanwhile, $\kappa/\kappa_{0}$ shows a plateau near the value of 2 at zero bias for lowest temperature of 1.2 K, but then the maximum value steadily decreases and the bias range where $\kappa/\kappa_{0}> 1$ narrows as temperature increases, ultimately reaches 1 for all biases at when $T > T_c$ at 10 K. Closely correlated changes in $\kappa/\kappa_{0}> 1$ and the magnitude of the superconducting gap independently confirm the detection of Andreev reflection. The reverse relationship can also be stated - observation of $\kappa/\kappa_{0}$ increasing to a maximum of 2 inside the gap identifies the origin of the gap as being most likely due to superconductivity. This is particularly important for a broad array of new emerging materials \cite{Liu2021,Zhao2021} as well as confined geometries of nanowires and nanoparticles \cite{Bezryadin2000}, where superconducting states are sought but competing ordered states, e.g. charge density wave, can manifest instead.

Theoretical modeling of NCAR was carried out with tight-binding simulations. 
From the theoretical point of view, we model our tunnel junction with the full Hamiltonian
$
\H = \H_{tip} + \H_{sample} + \H_{tunneling}
$,
where we consider an effective one-dimensional transport geometry
between the tip and the sample. We take a normal state tip
$
    \H_{tip} = 
    t_T \sum_{n\in \text{tip},s} c^\dagger_{n,s} c_{n+1,s}
    +
    h.c.
$
    a superconducting substrate
$
        \H_{sample} = 
    t_S \sum_{n\in \text{sample},s} d^\dagger_{n,s} d_{n+1,s}
        + \Delta d^\dagger_{n,\uparrow} d^\dagger_{n,\downarrow}
    + h.c.
$
    and a height-dependent tunneling coupling between the tip and the sample
$
        \H_{tunneling} = 
        \mathcal{V} (z)
    d^\dagger_0 c_0
    + h.c.
$
. $\mathcal{V} (z)$ parametrizes the distance-dependent tunneling amplitude between tip and
sample, that takes the form
$
    \mathcal{V} (z) = v_0 e^{-\lambda z}
$
with $\lambda$ the decay constant \footnote{We will take $t_S=t_T$, leading to a perfectly transparent regime for $v_0=t_T=t_S$ at $z=0$}. In the tunneling regime, leading-order perturbation theory yields a conductance
proportional to $dI/dV \propto \mathcal{V}^2$ for normal metal transport, and $dI/dV \propto \mathcal{V}^4$ for Andreev tunneling. The normalized decay rate is given by
$
    \rate = \kappa_{\Delta=\Delta}/\kappa_{\Delta=0}
$
where $\kappa_{\Delta=0}$ and $\kappa_{\Delta=\Delta}$ are the decay rates of the absence and presence of a superconducting gap, defined as
$
    \kappa = -\frac{1}{\lambda}\frac{dV}{dI}\frac{d^2 I}{dz dV}
$
evaluated at a bias $V$ and distance to the sample $z$.
We compute the $dI/dV$ and $\kappa$ factor using the non-perturbative $\mathcal{S}$-matrix formalism implemented with finite temperature non-equilibrium Nambu Green's functions \cite{Datta1995,Blonder1982,Sancho1985,PhysRevB.23.6851,PhysRevB.97.195429,pyqula}. Within this formalism, the differential conductance is computed exactly both in the tunneling and contact regimes at finite temperature.

Our theoretical calculations shown in Fig. \ref{fig2}(c) and \ref{fig2}(d) show that the temperature dependence accounts for both the smearing of the dI/dV spectra and the thermal renormalization of the decay rate. At zero temperature and inside the superconducting gap, the current is carried by Andreev reflection, involving two-tunneling events with $\kappa/\kappa_{0} = 2$. In the normal state, the current is carried by single charge tunneling with $\kappa/\kappa_{0} = 1$. At finite temperatures and inside the superconducting gap, both tunneling processes compete, giving rise to $\kappa/\kappa_{0}$ between 1 and 2. Inside the gap, and as the temperature is lowered, Andreev reflection becomes increasingly dominating, reaching the value of 2 for low enough temperatures. The transition can be parametrized by plotting maximum $\kappa/\kappa_{0}$ as a function of the ratio between the superconducting gap and temperature $\Delta / k_{B}T$ [Inset of Fig. \ref{fig:fig2}(b)]. The experimental points in the plot displays a very good match to the theoretical curve, which indicates among other things that the tight-binding model properly reflects the tunneling process of NCAR.

So far, our measurements and calculations point to a maximum value of $\kappa/\kappa_{0}$ of 2 in the tunneling limit\cite{Cuevas1996,Cuevas2003,Johansson2003}. Interestingly, at near-contact tunneling conductance we have also measured values \textit{in excess} of 2 [Fig. \ref{fig:fig3}(a)], whose origin reflects a much richer physical scenario in comparison with the tunneling regime. To address this question systematically, we have applied the normalized decay rate methodology to the near-contact regime, where tunneling conductance begins to approach quantum conductance $G_{0} = 2e^2/h$, as shown in Fig. \ref{fig3}. In contrast to the tunneling window in Fig. \ref{fig:fig1}(f), the $\rate$ spectra undergo a dramatic change as tip reaches near contact [Fig. \ref{fig3}(a,d)]. The center of plateau far exceeds the maximum value of $\rate = 2$ for the tunneling regime, reaching up to $\rate = 2.7$. Meanwhile, at the edges of the superconducting gap, the $\rate$ \textit{reduces} below 1, implying that the transmission is becoming more favorable than in the normal regime. Then, as the tip is brought even closer, the normalized decay rate in the middle of the superconducting gap begins to reduce. All these properties signify a distinct regime of Andreev reflection, that is intermediate between NCAR and PCAR (which we refer to as NeCAR below), corresponding to simultaneous enhancement and suppression of Andreev reflection and its strong energy dependence.

We now elaborate on the physical interpretation of the normalized decay rate in near contact regime. From a perturbation theory point of view, the decay rate directly signals the order of the term in perturbation theory dominating the current. In the tunneling limit of the normal state the leading contribution comes from second order in the coupling constant, whereas in the tunneling limit of Andreev reflection it comes from fourth order in the coupling constant, leading to a normalized ratio $\rate =2$. For near contact regime, besides the lowest order terms in perturbation theory higher order terms lead to competing contributions in the current \cite{PhysRevB.70.174509,Falci2001,Bignon2004}, such as sixth order in the coupling, leading to a normalized ratio $\rate >2$. As a result, the normalized decay rate is a direct probe of the order of the process responsible for the tunneling current as well as the nature of the current carriers in the substrate.

Furthermore, the enhancement of tunneling at the superconducting gap edge, which is marked by reduced $\kappa/\kappa_{0} < 1$, is the signature of resonant tunneling to the superconducting gap edge\cite{Datta1995}. The existence of a resonant process is directly accounted by the quantum transport formalism, and our experiment demonstrates its first unequivocal detection by NeCAR. Although in principle resonant tunneling should also be observed in the tunneling regime [Fig. \ref{fig:fig3}(b,e)], the resonance is extremely sharp and disappears when there is finite thermal broadening [Fig. \ref{fig:fig3}(c,f)]. However, around near contact, resonant tunneling starts to happen for broader bias range and it becomes detectable with spectroscopy. Tight-binding simulation at finite temperature reproduced all experimental features we observed at NeCAR [Fig. \ref{fig:fig3}(c,f)], showing that a phenomenology is fully captured by our theoretical model.

\begin{figure}[t!]
\center
\includegraphics[width=\linewidth]{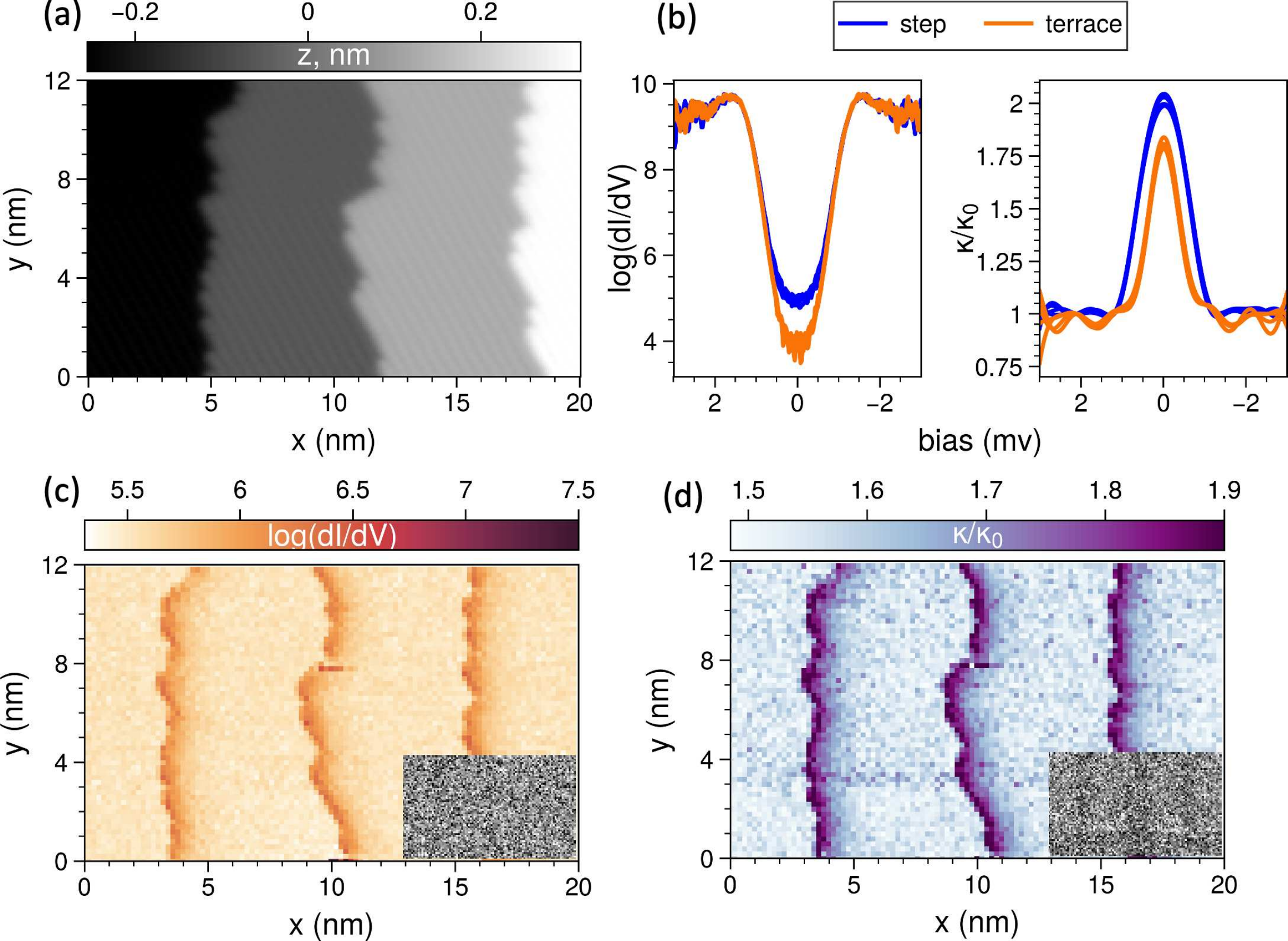}
\caption{Mapping Andreev reflection across intrinsic defects of Pb(110) crystal. (a) Constant current STM topography of with single atomic steps. (b) Comparison of $dI/dV$ (left) and $\rate$ (right) between steps and terraces. (c) The spatial distribution of the zero bias $dI/dV$ across the steps and terraces and a matching spatial distribution of $\rate$ at zero bias (d). Insets in (c) and (d) correspond to the maps at bias of 2.8 mV, -2.9 mV, respectively, both well outside the superconducting gap, where hardly any contrast in either observable is seen. \label{fig:fig4}}
\end{figure}

Clearly, the compatibility of NCAR and NeCAR spectroscopies with STM junction enables microscopic measurements in both regimes down to atomic resolution.  Intriguingly, NCAR revealed a new signatures of single atomic steps even on well-established Pb superconductor. Figure \ref{fig:fig4} shows the results of spectroscopy grid over an area of Pb(110) with single atomic steps shown in Fig. \ref{fig:fig4}(a). The steps show distinct contrast in both $\log(dI/dV)$ and $\kappa/\kappa_{0}$. Tunneling conductance is mainly seen to be enhanced in the narrow region around the gap center [Fig. \ref{fig:fig4}(b,c)], although the estimation of the size of the superconducting gap from the coherence peak position results in the same value for terrace and step within the measurement error [Fig. \ref{fig:fig4}(b)] and also as confirmed by Dynes fit \cite{Dynes1978}. Meanwhile, $\kappa/\kappa_{0}$ spectroscopy reveals that this difference is related not to single-charge density of states but to Andreev reflection, which is noticeably enhanced at the single atomic steps compared to surrounding terraces [Fig. \ref{fig:fig3}(b,d)]. The result indicates that single-charge DOS is nearly identical between the steps and terraces while Andreev reflection is significantly enhanced at the steps. In fact, from the perspective of Andreev reflection, the steps behave as though the superconductivity is locally enhanced, by comparison with the temperature effect in Fig. \ref{fig:fig2}(b). The origin of this enhancement remains to be understood. We hypothesize that it may be related to the reduced dimensionality of the step edge, and the concomitant effect on the local electronic structure sensitively probed through Andreev reflection.

To summarize, we have put forward NCAR as a powerful method for characterizing superconducting properties on nanoscale. Our methodology detects Andreev reflection by via a connection between the decay rate of the tunneling current and the order of the term in perturbation theory describing the current mechanism. We showed that NCAR can unequivocally distinguish a superconducting gap from other types of single-particle gaps. Furthermore, at near contact our methodology detects next to leading order Andreev reflection processes, which is distinct from both contact and tunneling regimes. Even in conventional s-wave Pb superconductor, mapping NCAR signal exhibited enhancement of Andreev reflection at single atomic steps, providing a powerful method to image current carrying excitations at the atomic scale. In the future, this technique can be readily extended to the emerging families of unconventional and topological superconductors \cite{Liu2021,Zhao2021,Dubouchet2018,Oh2021}, providing new and complementary insight into superconducting states and their localized excitations. Ultimately, our technique can allow probing the nature of current-carrying unconventional excitations in correlated quantum materials, ranging from Yu-Shiba-Rusinov states \cite{Huang2020,Peters2020} to fractionalized many-body excitations \cite{Zhu2020,Hashisaka2021}.

\textbf{Acknowledgements}:
Experimental measurements were supported by the U.S. Department of Energy, Office of Science, Materials Sciences and Engineering Division (W.K., P.M.). Experiments were carried out as part of the user project at the Center for Nanophase Materials Sciences, Oak Ridge National Laboratory, which is a US Department of Energy Office of Science User Facility. J.L.L. acknowledges the computational resources provided by the Aalto Science-IT project, and the financial support from the Academy of Finland Projects No. 331342 and No. 336243, and the Jane and Aatos Erkko Foundation.

\bibliography{biblio}

\end{document}